# SPH simulations of tidally unstable accretion disks in cataclysmic variables

J. R. Murray
*Canadian Institute for Theoretical Astrophysics*
*University of Toronto, Ontario, ON M5S 1A7*
*Canada*

## ABSTRACT

We numerically study the precessing disk model for superhump in the SU UMa subclass of cataclysmic variables, using a two dimensional SPH code specifically designed for thin disk problems.

Two disk simulations for a binary with mass ratio $q = \frac{3}{17}$ (similar to OY Car) are performed, in order to investigate the Lubow (1991 a,b) tidal resonance instability mechanism. In the first calculation, a disk evolves under steady mass transfer from $L_1$. In the second simulation, mass is added in Keplerian orbit to the inner disk. The two disks follow similar evolutionary paths. However the $L_1$ stream-disk interaction is found to slow the disk's radial expansion and to circularise gas orbits.

The initial eccentricity growth in our simulations is exponential at a rate slightly less than predicted by Lubow (1991a). We do not observe the clearing of material from the resonance region, via the disk's tidal response to the $m = 2$ component of the binary potential, described in Lubow (1992). Instead the $m = 2$ response weakens as the disk eccentricity increases.

Both disks reach an eccentric equilibrium state in which they undergo prograde precession. The rate of viscous energy dissipation in the disks has a periodic excess with a period matching the disks' rotation. The source is a large region in the outer disk. The mechanism by which the excess is produced is identified. The time taken for the periodic excess to develop is consistent with the first appearance of superhumps in a superoutburst.

**Key words:** accretion, accretion discs – hydrodynamics – instabilities – methods: numerical – binaries: close – novae, cataclysmic variables

## 1 INTRODUCTION

Dwarf novae are a category of cataclysmic variables that exhibit semi-periodic *outbursts* during which the system brightness increases by $2 - 5$ magnitudes. Cataclysmic variables are semi-detached binaries composed of a white dwarf (the detached component), and a low mass secondary star (the contact component). For a concise introduction to cataclysmic variables see Ritter (1992). In general, the secondary stars in cataclysmic variables are losing mass to the white dwarf via Roche lobe overflow. In the absence of a strong magnetic field, gas from the secondary forms a disk, centred upon the white dwarf. The source of dwarf novae outbursts is known to be this *accretion disk*.

SU UMa systems are a subclass of dwarf novae that also occasionally undergo *superoutbursts*, which are brighter and longer lasting than standard outbursts. The lightcurve of an SU UMa in superoutburst is unique in that a component of the signal is modulated with a period a few percent longer than the binary's orbital period. This modulation is known as *superhump*. For recent superhump observations the reader is referred to Hessman et al. (1992), and Molnar and Kobulnicky (1992).

The superhump signal is known to come from an extended region in the outer part of the accretion disk (Naylor et al. 1987). One possible explanation, dubbed the tidal resonance instability or TRI model by Lubow (1994), is that superhump is just the signature of an eccentric disk being periodically stressed by the binary's rotating tidal field. A slow prograde motion of the disk in the inertial frame is then responsible for the superhump period slightly exceeding the binary orbital period. In the TRI model the eccentricity is caused by a $3:1$ resonance between orbits in the disk and the orbit of the binary itself.

In this paper we investigate the TRI model by performing SPH simulations of disk evolution in close binary sys-tems. Firstly we review the model in more detail, contrasting the various simulations that were important in the model's development. Then we describe the adaptations needed to apply SPH to non-self gravitating thin disk problems. Of particular importance is the mechanism used to dissipate energy in the disk. We detail the artificial viscosity used in the code and provide an analytic estimate of the effective kinematic viscosity. This is then tested with axisymmetric ring spreading calculations. As a further test the result of a nonaxisymmetric simulation is compared against steady state thin disk theory. We then describe simulations of disk formation in the SU UMa system OY Car.

## 2 PREAMBLE

The first particle based simulations of disk formation in close binary systems were carried out by Lin and Pringle (1976). Reasoning that these accretion disks were geometrically thin, they restricted themselves to two dimensional simulations in the plane of the binary. Pressure forces in the disk were neglected. 'Viscous' forces were important and could not be ignored, but the details of the mechanism that produces viscosity in these disks was not known. Thus Lin and Pringle employed an ad hoc technique for removing energy from the disk. At regular time intervals $\Delta t$ a grid of mesh length $l$ was placed over the particles in the simulation. For each cell the particle velocities were adjusted to place them in rigid rotation about their (cell) centre of mass. This was done in such a way as to conserve linear and angular momentum at the cell level.

The gravitational influence of the disk upon the binary was neglected, as was disk self-gravity. Thus in between 'viscous' interactions, the particles moved independently of one another in the potential of the binary, as test masses in the restricted three body problem.

Hirose and Osaki (1990) pointed out that such a *sticky particle* scheme would have an effective kinematic viscosity

$$\nu_{\rm sp} = C\frac{l^2}{\Delta t}, \qquad (1)$$

where $C$ was a dimensionless constant to be determined empirically.

In this paper we use the same scalings as Lin and Pringle; length is scaled to the binary interstellar separation $d$, mass is scaled to the total system mass $M = M_{\rm s} + M_{\rm p}$, and time is scaled to the reciprocal of the binary's orbital angular frequency, ${\Omega_{\rm b}}^{-1}$. In these units a time interval of $2\pi$ corresponds to one period of the binary. The geometry of the system is defined by the mass ratio

$$q = \frac{M_{\rm p}}{M_{\rm s}}. \qquad (2)$$

Note that this definition of $q$ agrees with most authors but is the inverse of the definition in Lin and Pringle.

Although the Lin and Pringle calculations were perhaps hampered by insufficient resolution, they provided a blueprint for many subsequent simulations. Lin and Pringle built up their disks from zero initial mass, introducing particles at a steady rate from the inner Lagrangian ($L_1$) point. Lubow and Shu (1975) had previously calculated that gas overflowing the secondary Roche lobe would fall into the primary Roche lobe in a narrow stream and calculated its velocity and direction as a function of the binary mass ratio.

As an inner boundary condition for the disk, Lin and Pringle simply used a hole in the domain. Any particle ending a time step less than a distance $r_{\rm wd}$ from the primary was deleted from the calculation. Similarly any particles that returned to the secondary Roche lobe or that found themselves a very large distance from the primary were removed and recorded.

Lin and Pringle performed simulations for three mass ratios, $q = 0.4, 1.0, 2.5$. In each case they found the disk reached an equilibrium state where the rate of infall from $L_1$ balanced the accretion onto the primary. The total angular momentum of the disk also reached a stable value. It was noted that very few particles 'escaped' the white dwarf Roche lobe. This implied that tidal torques were the principal means of removal of angular momentum from the outer edge of the disk.

Whitehurst (1988b) implemented a fully Lagrangian particle method for modelling accretion disks. In contrast to Lin and Pringle he introduced dissipation into his code by allowing particles to undergo inelastic collisions with one another. Whitehurst also allowed for a repulsive 'pressure' force between particles. No grid was required to calculate these forces.

The boundary conditions used by Whitehurst (particle removal at the white dwarf surface etc.) were similar to Lin and Pringle's. Whitehurst chose to make the mass input rate from the secondary vary with time. Pressure forces had been included so an equation of state was required. Whitehurst assumed that all the heat generated by viscous action in the disk was instantaneously radiated away. Thus he set the speed of sound to a constant value.

Whitehurst (1988b) described a calculation for a binary with $q = \frac{8}{13}$. The disk developed a two armed spiral structure and was elongated perpendicular to the binary axis. The disk appeared stationary in the binary frame. The mean outer radius of this disk corresponded with Paczyński's (1977) estimate. Paczyński had assumed that gas streamlines in the disk would be well approximated by a family of simply periodic orbits of noninteracting particles in the restricted three body problem. Close to the primary these orbits are approximately circular and concentric. At larger distances the orbits are more distorted and at larger radii again the orbits begin to intersect one another. Paczyński proposed that the disk would be truncated at the last nonintersecting simply periodic orbit of a noninteracting particle (LNO). This is referred to as *tidal truncation* of the disk.

A simulation of a system with a low mass secondary (q=0.15, Whitehurst 1988a) produced a more eccentric disk. In the binary frame the disk's semi-major axis rotated in the opposite sense to the motion of the disk gas with a period slightly exceeding the binary period. (In the inertial frame the disk exhibited a slow prograde precession). The energy dissipation in the disk had a cyclic component corresponding to the disk rotation period. Whitehurst proposed that superhump in SU UMa binaries was due to the same tidal instability seen in his simulation. The superhump period correlated with the rotation of the simulation disk. Also, the tidal instability only occurred in low mass ratio systems, and SU UMa systems characteristically have low mass ratios.

Hirose and Osaki (1990) performed a series of two di-

mensional simulations for different mass ratios using the Lin and Pringle sticky particle method. In each calculation they allowed a disk to form from scratch via *steady* mass transfer from the secondary. They found that in general, disks in low mass ratio systems were tidally unstable whilst those in high mass ratio systems were stable.

The range of mass ratios for which tidal instability occurred depended upon the resolution (and effective viscosity) of their simulations. One series of simulations with grid size $l = 0.01\,d$ (using the same scalings as above) produced stationary disks for mass ratios $q > 0.25$. A second set of calculations with $l = 0.02\,d$, and hence with four times the effective viscosity, gave forth tidally unstable disks in all but the $q = 1$ case. In an argument that built upon Whitehurst's hypothesis, Hirose and Osaki proposed that the tidal instability was caused by a 3 : 1 resonance between orbits in the disk and the orbit of the binary itself. With an analysis of the orbits of noninteracting particles they found that the 3 : 1 resonance only occurred inside the tidal radius (LNO) for systems with $q \lesssim 0.25$. Hirose and Osaki rationalised the differing stability results of the two simulation sets by explaining that in the $l = 0.02\,d$ calculations, the large effective viscosity caused the disks to spread significantly beyond the tidal radius, allowing them to interact with the 3 : 1 resonance in larger mass ratio systems.

Lubow (1991a) added a good deal to the sophistication of the model. Rather than considering the orbits of noninteracting particles, he completed an analysis of a fluid disk or ring's response to a tidal field. Knowing that eccentric Lindblad resonances cause eccentricity growth, he identified the $m = 3$ eccentric inner Lindblad resonance as being the most important resonance for increasing eccentricity in accretion disks in close binary systems. With a nonlinear mode analysis Lubow went on to identify the feedback mechanism causing the eccentricity to grow and then calculated a growth rate.

In a set of cylindrical coordinates $(r, \theta)$ centred upon the primary, the disk eccentricity can be expressed as a sinusoidal mode with argument $\theta$. The *tidal field* due to the secondary star can be decomposed into a set of functions $\phi_m(r) cos[m(\theta - \Omega_{\rm b} t)]$, where $m$ is an integer. The $m = 3$ component of the tidal field generates a sinusoidal response in the disk with argument $3\theta - 3\Omega_{\rm b} t$. At the Lindblad resonance this response couples with any existing eccentricity to produce a travelling wave (argument $2\theta - 3\Omega_{\rm b} t$). In turn, the interaction of the wave and the tidal response acts to reinforce the disk's eccentricity, which grows exponentially with a rate

$$\lambda = 2.08 \times 2\pi\, \Omega_{\rm b}\, q^2\, {r_{\rm res}}^2\, \frac{\Sigma(r_{\rm res})\, e(r_{\rm res})}{M_{\rm d}\, \langle e \rangle}. \qquad (3)$$

Here $r_{\rm res}$ is the radius of the Lindblad resonance, $\Sigma(r)$ is the surface density at radius $r$, $M_{\rm d}$ is the total disk mass, $e(r)$ is the eccentricity at radius $r$, and $\langle e \rangle$ is the mass averaged disk eccentricity.

In a second paper, Lubow (1991b) detailed several SPH calculations that confirmed both the mechanism and the calculated growth rate. His simulations however differed from those of Whitehurst, and from those of Hirose and Osaki. Rather than allowing a disk to build up via mass transfer from the secondary, Lubow prescribed an initial disk with particles laid down on the simple periodic orbits of noninteracting particles. The mass transfer stream was completely absent from Lubow's calculations. Secondly, in order to view the newly discovered mechanism in isolation, Lubow initially included only the $m = 3$ component of the tidal field in his calculations. He found that when the full binary potential was included, the eccentricity growth rate was reduced. He later determined (Lubow 1992) that non-resonant tidal stresses due to the $m = 2$ component of the tidal field repelled material from the inner Lindblad resonance and hence reduced the instability's strength. Lubow expressed concern that the weakened instability would be incapable of producing a superhump response sufficiently rapidly after the onset of superoutburst.

The major factors contributing to the disk's precession (a slow prograde motion with respect to the inertial frame that, in the TRI model, is responsible for the superhump period slightly exceeding the binary period) were detailed in Lubow (1992). The $m = 0$ component of the tidal field induces a prograde precession by shifting the epicyclic frequency from the Keplerian value, $\Omega_{\rm b}$. Secondly, Lubow found that pressure waves would produce a retrograde precession. A third factor, due to wave stresses, was found to be capable of producing both a prograde or a retrograde procession.

The analysis up till that point did not consider the possible effects of a sudden, dramatic increase in viscosity in the disk that would occur in the thermal instability model of a dwarf novae outburst. Neither did the model account for influence of the gas stream from the secondary. Lubow (1994) did examine the rôle the stream might play in creating or increasing eccentricity in the disk. In particular he considered the periodic variation in the momentum the gas stream would impart to a precessing elliptical disk and asked whether it might act to increase the eccentricity. He found in fact, that it would reduce the disk's eccentricity. The analysis did not consider the distortions upon disk streamlines that the secondary would impart.

There is a second question to be asked with regard to the gas stream. We know that, in comparison to gas in the outer disk, the gas stream has relatively low angular momentum and their collision will reduce specific angular momentum in the outer disk. So does the stream disk collision act to push gas into the resonant region? We will try to answer this question in this paper.

To reiterate, disagreement over the Tidal Resonance Instability (TRI) model for superhump appears to be focused upon whether the gas is capable of reaching the 3 : 1 resonance. If that issue is to be resolved then we must know more about the mechanism dissipating energy and transporting angular momentum through the disk.

To emphasise the previous point we consider the simulations of Heemskerk (1994). He assumed the gas to be inviscid, and conducted simulations on that basis using an Eulerian code. In calculations he performed that included only the $m = 3$ component of the secondary's potential, he found the eccentricity grew in agreement with Lubow's model. However, when Heemskerk included the full binary potential in his simulations, he found the disk to be truncated somewhat inward of the resonance region and that there was no eccentricity growth. But, apart from any numerical viscosity present in his code, Heemskerk had not explicitly included a mechanism for transporting angular momentum through the disk. It is not surprising therefore that his disks should have a smaller outer dimension than the highly viscous disk calculations of say Hirose and Osaki. As an aside we remark that Heemskerk was incorrect in referring to the calculations of Whitehurst and of Hirose and Osaki as being SPH simulations. The *sticky particle* method used by Lin and Pringle, and Hirose and Osaki relies upon a grid for determining the energy dissipation. SPH on the other hand is characterised by its independence of any mesh scheme. And whilst Whitehurst's unique particle scheme was also fully Lagrangian, it did not explicitly include the kernel interpolation scheme that is central to SPH.

## 3 NUMERICAL METHOD

The numerical simulations in this paper were completed using Smoothed Particle Hydrodynamics (SPH), a fully Lagrangian particle method. For a comprehuensive review of SPH, the reader is referred to Monaghan (1992). In this section we discuss the specific application of the SPH technique to accretion disk problems. First, an appropriate set of SPH equations for geometrically thin disks is obtained. Then dissipation in the SPH scheme is discussed. The standard artificial viscosity term is modified to a form more suitable to the modelling of viscous shear, and a theoretical estimate of its equivalent kinematic viscosity is obtained. The estimate is tested in the next section. The artificial viscosity is related to the Shakura-Sunyaev viscosity parameterisation commonly used in thin disk theory. Finally we consider the fact that viscous time scales in an accretion disk are typically much longer than dynamical time scales, and discuss the problems this presents for numerical modelling.

### 3.1 SPH Equations for a Geometrically Thin Disk

We start from the basic equation for an integral interpolant,

$$A_{\rm I}(\mathbf{r}) = \int A(\mathbf{r}')\, W(\mathbf{r}-\mathbf{r}', h)\, \mathbf{dr}'. \qquad (4)$$

$W(\mathbf{r}, h)$ is an interpolating *kernel*, and $h$ is the smoothing length. The functions $A$ that we are interested in describe disk properties that have in some sense been vertically integrated to account for the disk thickness, and so are functions defined on the disk midplane. Accordingly we will use cylindrical coordinates $(\mathbf{r}(r,\theta), z)$ with $z = 0$ on the midplane. An element of disk surface area $\mathbf{dr}$ with surface density $\Sigma$ has mass

$$m = \frac{\mathbf{dr}}{\Sigma}. \qquad (5)$$

As usual in SPH the integral interpolant is approximated by a summation over a set of disordered particles,

$$A_s(\mathbf{r}) = \sum_b m_b \frac{A_b}{\Sigma_b} W(\mathbf{r}-\mathbf{r}_b, h), \qquad (6)$$

with $m_b$ being the mass of particle $b$. Thus the surface density at $\mathbf{r}$ is given by

$$\Sigma_s(\mathbf{r}) = \sum_b m_b W(\mathbf{r}-\mathbf{r}_b, h). \qquad (7)$$

To (approximately) find the volume density $\rho$ on the midplane we divide $\Sigma$ by the disk's pressure scale height, which is given by

$$H = \frac{c}{\Omega_{\rm b}}. \qquad (8)$$

We now want the SPH equivalents for the momentum and energy equations for thin disks. For particle $a$, the standard SPH momentum equation incorporating artificial viscosity is

$$\begin{aligned}\frac{\mathrm{d}\mathbf{v}_a}{\mathrm{d}t} = & - \sum_b m_b \Big(\frac{P_a}{\rho_a^2} + \frac{P_b}{\rho_b^2} \\ & + \frac{\beta {\mu_{ab}}^2 - \zeta \bar{c}_{ab}\mu_{ab}}{\bar{\rho}_{ab}}\Big) W(\mathbf{r}_a - \mathbf{r}_b, h). \qquad (9)\end{aligned}$$

Here $\bar{c}_{ab} = 0.5(c_a + c_b)$ is the average sound speed for particles $a$ and $b$. $\bar{\rho}_{ab}$ is similarly defined. The notation is made more concise by writing $W(\mathbf{r}_a - \mathbf{r}_b, h)$ as $W_{ab}$. $\zeta$ and $\beta$ are the linear and nonlinear artificial viscosity parameters respectively.

$$\mu_{ab} = \begin{cases} \frac{\mathbf{v}_{ab}\cdot\mathbf{r}_{ab}}{{r_{ab}}^2 + \eta^2} & \mathbf{v}_{ab}\cdot\mathbf{r}_{ab} \le 0 \\ 0 & \mathbf{v}_{ab}\cdot\mathbf{r}_{ab} > 0 \end{cases}, \qquad (10)$$

where $\mathbf{v}_{ab} = \mathbf{v}_a - \mathbf{v}_b$ and $\mathbf{r}_{ab} = \mathbf{r}_a - \mathbf{r}_b$. $\eta$ is a constant with the dimensions of length. Typically $\eta = 0.1\,h$.

We take the $(r,\theta)$ component of the fluid momentum equation,

$$\frac{\mathrm{d}\mathbf{v}_{r,\theta}}{\mathrm{d}t} = \cdots + \frac{1}{\rho}\nabla_{r,\theta} P. \qquad (11)$$

Following Goldreich and Lynden-Bell (1965) we multiply Eqn 11 by $\rho$ and integrate over $z$ to obtain

$$\int_{-\infty}^{\infty} \rho \frac{\mathrm{d}\mathbf{v}_{r,\theta}}{\mathrm{d}t}\,\mathrm{d}z = \cdots + \int_{-\infty}^{\infty} \nabla_{r,\theta} P \,\mathrm{d}z. \qquad (12)$$

Reversing the order of operations on the right hand side, and assuming $\mathbf{v}_{r,\theta}$ is independent of $z$, we get

$$\begin{aligned}\frac{\mathrm{d}\mathbf{v}_{r,\theta}}{\mathrm{d}t} &= \cdots + \frac{1}{\Sigma}\nabla_{r,\theta}\int_{-\infty}^{\infty} P\,\mathrm{d}z \qquad (13)\\ &\simeq \frac{1}{\Sigma}\nabla_{r,\theta}(PH). \qquad (14)\end{aligned}$$

We rewrite the pressure gradient term,

$$\frac{1}{\Sigma}\nabla_{r,\theta}(PH) = \nabla_{r,\theta}\Big(\frac{PH}{\Sigma}\Big) + \frac{PH}{\Sigma^2}\nabla_{r,\theta}\Sigma. \qquad (15)$$

Ignoring the artificial viscosity term for a moment, our two dimensional SPH momentum equation for a geometrically thin disk is

$$\frac{\mathrm{d}\mathbf{v}_a}{\mathrm{d}t} = -\sum_b m_b \Big(\frac{P_a H_a}{\Sigma_a^2} + \frac{P_b H_b}{\Sigma_b^2}\Big)\nabla_a W_{ab}. \qquad (16)$$

Similarly, our SPH energy equation is

$$\frac{\mathrm{d}u_a}{\mathrm{d}t} = \frac{1}{2}\sum_b m_b \Big(\frac{P_a H_a}{\Sigma_a^2} + \frac{P_b H_b}{\Sigma_b^2}\Big)\mathbf{v}_{ab}\cdot\nabla_a W_{ab}. \qquad (17)$$

### 3.2 SPH Artificial Viscosity

The standard SPH artificial viscosity expression, as seen in Monaghan, is composed of a linear term and a quadratic term. The quadratic ($\beta$) term, which is similar to a Von Neumann-Richtmyer viscosity, was originally included to improve the treatment of high Mach number shocks. We are interested in a supersonic shearing flow for which the linear ($\zeta$) term alone is sufficient. Thus we have set $\beta = 0$, as did Meglicki et al. (1993). Traditionally $\alpha$ is used to denote the SPH linear artificial viscosity parameter. We have changed the notation to avoid confusion with the Shakura-Sunyaev parameter.

In the standard formulation, the artificial viscosity term is set to zero for expanding flows. Again we realise that the viscous term was originally intended to improve SPH's handling of compression, and dissipation in regions of rarefaction was undesirable. Here however, with a strongly shearing flow, it makes no sense to use such a *viscous switch*. The remaining linear $\zeta$ term generates both shear and bulk viscosity.

By reversing the procedure used to generate the SPH summation equations we can obtain a continuum limit for this artificial viscosity term (see e.g. Pongracic (1988), or Murray (1994)). When we let the number of particles $N \rightarrow \infty$ and the smoothing length $h \rightarrow 0$ we obtain

$$\mathbf{a}_{\mathrm{v}} = \frac{\zeta h \kappa c}{2\Sigma} \left( \nabla \cdot (c\Sigma \mathbf{S}) + \nabla (c \Sigma \nabla \cdot \mathbf{v}) \right), \tag{18}$$

where the deformation tensor

$$S_{ab} = \frac{\partial v_a}{\partial x_b} + \frac{\partial v_b}{\partial x_a}. \tag{19}$$

$\kappa$ is a constant determined by the interpolation kernel being used. In two dimensions, with the cubic spline kernel given in Monaghan (1992), we find

$$\kappa = \frac{1}{4}. \tag{20}$$

If we assume the density and sound speed vary on length scales much longer than the velocity we have

$$\mathbf{a}_{\mathrm{v}} = \frac{\zeta h c}{8} \left( \nabla^2 \mathbf{v} + 2\nabla(\nabla \cdot \mathbf{v}) \right). \tag{21}$$

If we assume the velocity divergence is zero then we have

$$\mathbf{a}_{\mathrm{v}} = \frac{\zeta h c}{8} \nabla^2 \mathbf{v}. \tag{22}$$

Thus in the continuum limit we could expect the linear artificial viscosity term to generate a shear viscosity

$$\nu = \frac{1}{8} \zeta c h. \tag{23}$$

The form of Eqn 23 is similar to the Shakura Sunyaev viscosity parametrisation

$$\nu = \alpha c H. \tag{24}$$

In the SPH case however, the length scale over which dissipation occurs is the smoothing length, rather than the pressure scale height. A disk with kinematic viscosity given by Eqn 23 has a viscous dissipation equivalent to an $\alpha$ disk with

$$\alpha = \frac{1}{8} \left( \frac{q}{1+q} \right)^{\frac{1}{2}} \frac{\zeta}{c} r^{-\frac{3}{2}}. \tag{25}$$

**Figure 1.** Radial surface density plots at times $t = 0.0, 0.2, 0.4$ and $0.8$ for the axisymmetric ring spreading calculation. Parameters used in the simulation are listed in Table 1. Eqn 32 is also plotted (solid lines). For this simulation, particles were laid down in 21 concentric rings which were $\frac{1}{2}h$ apart.

**Figure 2.** Maximum surface density of the simulation depicted in Fig. 1, as a function of time (heavy dots). Eqn 32 evaluated at $r = r_0$ (solid line) is plotted for comparison.

The simulations described later in this paper have constant $c$, $\zeta$ and $h$. In terms of the Shakura-Sunyaev parameterisation, this is like setting $\alpha \propto r^{-\frac{3}{2}}$.

### 3.3 Axisymmetric Tests

In order to verify our analysis of the standard SPH artificial viscosity term we completed a series of axisymmetric ring spreading simulations for comparison with semianalytic solutions. Lynden-Bell and Pringle (1974) considered the evo-

**Table 1.** Parameter settings used in the axisymmetric ring spreading calculations

| Parameter | Value | Remark |
|---|---|---|
| $\zeta$ | 10.0 | SPH artificial viscosity parameter |
| $h$ | 0.01 | SPH smoothing length |
| $c$ | 0.02 | Sound speed |
| $r_0$ | 0.85 | centre of Gaussian |
| $r_1$ | 0.80 | inner edge of annulus |
| $r_2$ | 0.90 | outer edge of annulus |
| $l$ | 0.025 | half width of Gaussian |
| $\Delta r$ | $\frac{1}{2}h$ | initial interparticle spacing |

**Figure 3.** Maximum surface density of a second ring spreading simulation as a function of time (heavy dots). Parameters are identical to the previous simulation with the exception that $\Delta r = \frac{5}{13}\,h$. Again Eqn 32 evaluated at $r = r_0$ (solid line) is plotted for comparison.

lution of an axisymmetric ring with an initial density distribution

$$\Sigma_0(r) = \frac{1}{2\pi r_0}\,\delta(r - r_0), \tag{26}$$

where $\delta$ is the Dirac delta function and $r_0$ is the ring's initial radius. For a constant kinematic viscosity $\nu$, and a Keplerian potential, the solution for $t > 0$ (Eqn 2.13, Pringle 1981) is

$$\Sigma(r,t) = \frac{1}{\pi r_0^2 x^{\frac{1}{4}} \tau}\,\exp\left[-\frac{(1+x^2)}{\tau}\right] I_{\frac{1}{4}}\left(\frac{2x}{\tau}\right), \tag{27}$$

where $x = \frac{r}{r_o}$ and $\tau = \frac{12\nu t}{r_0^2}$ are dimensionless radius and time variables respectively, and $I_{\frac{1}{4}}$ is a modified Bessel function of the first kind. Now for large arguments

$$I_{\frac{1}{4}}(x) \simeq \frac{e^x}{\sqrt{2\pi x}} \qquad x \gg 1. \tag{28}$$

So for small $\tau$,

$$\Sigma(r,t) \simeq \frac{1}{2\pi^{\frac{3}{2}} r_0{}^2 x^{\frac{3}{4}} \tau^{\frac{1}{2}}}\exp\left[-\frac{(1-x)^2}{\tau}\right] \qquad \tau \ll 1. \tag{29}$$

The density distribution at some small time $\tau$ is thus approximately a Gaussian modified by factor $x^{-3/4}$.

Flebbe et al. (1994) sought to test their SPH code, which incorporated a tensor form of a general Navier Stokes viscosity term, by simulating such a ring. Naturally with SPH they could not set up the initial $\delta$ density distribution so they started the simulation at dimensionless time $\tau = 1.6 \times 10^{-4}$. It is not clear how Flebbe et al. constructed their ring to obtain the correct density profile. In general one has the option of either manipulating the particle number density in the ring or initialising the particles with different masses.

Rather than simulate the evolution of a $\delta$ function ring, we chose to make use of the fact that when $\nu$ is constant (or indeed when $\nu$ is a function of $r$ only) the equation for the evolution of an axisymmetric ring is linear in the density and a solution for a general initial condition can be obtained from the $\delta$ function response. Thus

$$\Sigma(r,t) = \frac{1}{6 r^{\frac{1}{4}} \nu t}$$

$$\int_0^\infty \Sigma_0(r') \quad \exp\left[-\tfrac{r^2 + r'^2}{12\nu t}\right] \quad I_{\frac{1}{4}}\left(\frac{r' r}{6\nu t}\right) \mathrm{d}r'. \tag{30}$$

Upon trying several initial density functions, we found that a smoothly varying $\Sigma_0$ gave best results. We thus chose a Gaussian,

$$\Sigma_0(r) = \begin{cases} \exp\{-\left(\frac{r-r_o}{l}\right)^2\} & r_1 < r < r_2 \\ 0 & \text{elsewhere} \end{cases}, \tag{31}$$

truncated at $r_1$ and $r_2$. $r_0$ and $l$ are constants specifying the radius of maximum density and the width of the Gaussian curve respectively. Using Eqn 28 we obtain, for small time $t$, the approximate solution

$$\Sigma(r,t) \approx \frac{1}{r^{3/4}\sqrt{\pi 12\nu t}}$$

$$\int_{r_1}^{r_2} r'^{\frac{3}{4}} \quad \exp\left[-\left(\tfrac{r' - r_0}{l}\right)^2\right] \quad \exp\left[-\frac{(r' - r)^2}{12\nu t}\right] \mathrm{d}r'. \tag{32}$$

Once the density is known, the radial component of the velocity is given by

$$v_r(r,t) = -3\nu\frac{\partial}{\partial r}\left[\ln(r^{\frac{1}{2}}\Sigma(r,t))\right]. \tag{33}$$

For $\Sigma_0$ given by Eqn 31 we have

$$v_r(r,0) = -\frac{3\nu}{2r} + \frac{6\nu}{l^2}(r - r_0). \tag{34}$$

In order to initialise our ring we require the kinematic viscosity. In fact we simply substituted Eqn 23 for $\nu$ in Eqn 34.

In the simulations described here, the kinematic viscosity was assumed to be constant. The code used a smoothing length $h$ that was constant in time and space. Pressure forces were switched off so we could study the artificial viscosity term in isolation. However, as $\nu \propto c$, the sound speed was not set to zero. Best simulation results were obtained when the initial particle distribution reflected the symmetry of the system. Particles were laid out in a series of concentric rings $\Delta r$ apart, so as to give a constant number density throughout the ring. Particle masses were initialised to give the Gaussian density profile. To provide comparison with simulation results, Eqn 32 was numerically integrated using MATHEMATICA.

In Fig. 1, simulation density profiles (heavy dots) at the times shown are compared with Eqn 32. The various parameters used in this simulation are summarised in Table 1. Units were chosen so that the mass of the central body, and the orbital angular frequency of a test particle at unit radius, were both equal to one. Flebbe et al. (1994) stated that in their simulation the radial velocity was less than 5% of the azimuthal velocity. In our case the figure is approximately 10%. Eqn 25 indicates that this simulation generates a dissipation comparable with a Shakura-Sunyaev viscosity with parameter $\alpha = 0.8$. This is not to say that the SPH artificial viscosity is producing a dissipation equivalent to a Shakura-Sunyaev term. We merely include the figure for order of magnitude comparison.

Densities in the simulation were calculated using Eqn 7, as opposed to integrating the continuity equation. As a result the initial peak density in the simulation is approximately 5% less than that of the Gaussian. This is because $h$ is comparable to the Gaussian half width $l$. At the peak the density gradient changes significantly over a smoothing length. Improved resolution could be gained by reducing $h$ or increasing $l$. A similar simulation with smaller $h$ would require an increased number of particles, and hence more computer time. On the other hand the ring spreads with a time scale that is proportional to $l^2$ so increasing the Gaussian half width would also lead to more time consuming simulations. Fig. 1 may be compared to Fig. 2 of Flebbe et al. Their results show more scatter about the analytical curves, possibly as a result of the way in which their particles were set up.

In Fig. 2 we plot the ring's peak density as a function of time (heavy dots). We also show $\Sigma(r_0, t)$, i.e. Eqn 32 evaluated at $r_0$. For small time $t$ this gives a good estimate of the ring's maximum density. At first glance the agreement between theory and simulation does not appear as good as Fig. 1 suggested. We have already discussed why the simulation peak density should be slightly reduced at small time. At times $t > 1$ the simulation density is larger than the theoretical value and appears to be levelling off more rapidly. As the annulus spreads, the gaps between the constituent rings of particles get larger until finally they are comparable with the smoothing length and the rings lose contact with one another. At that point, a particle's only near neighbours lie on the same ring. This is a resolution problem which can be overcome by increasing the number of particles or alternatively by incorporating a variable smoothing length. To confirm our explanation we repeated the simulation with an increased number of particles ($\Delta r$ was reduced from $\frac{1}{2}h$ to $\frac{5}{13}h$). Fig. 3 shows shows that the resulting peak density profile agrees much better with the semianalytic curve.

Artymowicz and Lubow (1994) used similar tests to estimate the shear viscosity of their SPH code. In contrast to our artificial viscosity term which is added to the pressure term in the momentum equation, Artymowicz and Lubow choose a multiplicative term. The nature of the term is essentially the same as ours, differing only by a factor $\frac{\Sigma}{c^2}$ to maintain dimensional consistency. Artymowicz and Lubow also found it appropriate to remove the *viscous switch*. Pongracic took a term similar to Artymowicz and Lubow's to the continuum limit, in the manner described in the previous section, and obtained an equation similar but not identical to Eqn 18. Once the appropriate assumptions have been made however, such a term would also produce the shear viscosity given by Eqn 23. Without the aid of such an analysis, Artymowicz and Lubow obtained via dimensional arguments the viscosity relation

$$\nu = q\,\zeta\,c\,\langle h\rangle, \tag{35}$$

with $q$ being a constant of proportionality obtained empirically from ring spreading calculations. As their code used a variable smoothing length, Artymowicz and Lubow used a locally averaged smoothing length $\langle h\rangle$ in Eqn 35. For simulations using a three dimensional code with the cubic spline kernel they found $q = 0.1$ to within 30%. This would appear to be more in agreement with the analysis for the two dimensional rather than the three dimensional code. However we note that because the Artymowicz and Lubow artificial viscosity term multiplies the pressure term in the momentum equation, they could not switch the pressure forces off to consider the effects of the viscous dissipation term in isolation. The pressure causes an additional spreading of the ring which leads to an error in estimating $q$. Artymowicz and Lubow do not go into the details of their tests, and in particular it is not clear how they chose their initial radial velocity condition.

**Figure 4.** Azimuthally averaged surface density plotted as a function of $r$ for the $q = 1$ simulation at $t = 1000.00\,\Omega_b^{-1}$ (heavy dots). Simulation parameters are listed in Table 2. Eqn 36 with $r_* = 0.017\,d, \nu = 2.35 \times 10^{-4}\,d^2\,\Omega_b^{-1}$ gives best fit to the data and is plotted as the solid line.

**Figure 5.** Azimuthally averaged rate of energy dissipation in the $q = 1$ simulation at $t = 1000.00\,\Omega_b^{-1}$ plotted on logarithmic axes (heavy dots). Eqn 38 with $r_* = 0.017\,d$ is also shown (solid line).

We found that if, instead of using the kinematic viscosity given by Eqn 23, we used a different $\nu$ to determine the ring's initial radial velocity, the agreement between simulation and theory deteriorated. Thus we were able to confirm that Eqn 23 gave an accurate estimation of the viscosity, to within an uncertanity of approximately 10%.

### 3.4 Nonaxisymmetric Tests

As discussed in Section 2, simulations by Lin and Pringle (1976), Whitehurst (1988b), and Hirose and Osaki (1990) for high mass ratio systems had resulted in steady state disks that were approximately axisymmetric and stationary in the binary frame. In fact Hirose and Osaki used such a simulation to determine the viscous dissipation of their *sticky particle* code.

The surface density in a steady state axisymmetric disk is given by

$$\nu\Sigma = \frac{\dot{M}}{3\pi}\left(1 - \sqrt{\frac{r_*}{r}}\right), \tag{36}$$

where $\dot{M}$ is the steady state mass flux through the disk and $r_*$ is the radius of the disk's inner edge. The inner boundary condition implicit in Eqn 36 is

$$\frac{{\rm d}\Omega(r_*)}{{\rm d}r} = 0. \tag{37}$$

**Table 2.** Parameters for $q = 1$ disk simulation.

| Parameter | Value | Remark |
|---|---|---|
| $\zeta$ | 10.0 | SPH artificial viscosity parameter |
| $h$ | $0.01\,d$ | SPH smoothing length |
| $c$ | $0.02\,d\,\Omega_{\rm b}^{-1}$ | Sound speed |
| $v_{\rm inj}$ | $0.1\,d\,\Omega_{\rm b}^{-1}$ | velocity of particles injected at $L_1$ |
| $\theta_{\rm inj}$ | 0.3491 rad | angle to binary axis of injected particles |
| $\Delta t$ | $0.01\,\Omega_{\rm b}^{-1}$ | interval between particle injection |
| $r_{\rm wd}$ | $0.02\,d$ | white dwarf radius |

Hirose and Osaki took the steady final state of their simulation and fit the density distribution to Eqn 36, using $r_*$ and $\nu$ as free parameters.

The rate at which energy is dissipated per unit area in a steady state axisymmetric disk

$$D(r) = \frac{3\dot{M}}{8\pi r^3}\left(1 - \sqrt{\frac{r_*}{r}}\right). \tag{38}$$

Hirose and Osaki compared energy dissipation in their simulation with Eqn 38. In this case the single free parameter is $r_*$. The dissipation comparison thus provided a check upon the density comparison.

For us, the nonaxisymmetric test provides a useful intermediate step between the ring spreading calculations and the simulations we are interested in finally performing. Full disk simulations typically take several tens of thousands of time steps and there is no direct way of measuring their accuracy or the validity of the result. With the particle number constantly changing and viscous forces at work, there are no *constants of the motion* that can be monitored as the simulation proceeds. In contrast the ring spreading calculations could be completed in a hundred or so time steps and we were able to follow monitor conservation of angular momentum. We briefly describe therefore a simulation in a system with mass ratio $q = 1$.

We use similar boundary conditions to Lin and Pringle (1976). In particular we choose the simplest inner boundary condition, that is a circular hole in the domain of radius $r_{\rm wd}$ centred upon the primary. Any particle ending a time step within this circle is simply dropped from the calculation. We record and remove any particles that return to the secondary or that are flung to a large radius.

Initially, simulations were run with the pressure forces set to zero, the only interparticle forces being due to the artificial viscosity term. It was found however that a number of tightly bound particle pairs formed. The artificial viscosity force between two particles always opposes the relative motion. Thus it was possible for two particles to approach each other very closely and become 'stuck' to one another. The fraction of particles that formed such pairs was only ever very small (less than half a percent). But they produced density extremes which caused the Courant condition time step to become very small, and hence the simulation to become very slow. However pressure forces, which are always repulsive, act to prevent the formation of such pairs. Thus the pressure terms were included in all further simulations.

The simulation began with no particles in the disk. Then, following Hirose and Osaki, single particles were injected at regular intervals $\Delta t$ from the inner Lagrangian point. Simulation parameters are summarised in Table 2. For these values Eqn 23 predicts $\nu = 2.5 \times 10^{-4}\,d^2\,\Omega_{\rm b}$.

We followed the calculation to $t = 1000.00\,{\Omega_{\rm b}}^{-1}$ (159 binary periods). The disk reached a steady state after about 100 binary periods.

The radial density profile for the simulation disk at $t = 1000.00\,{\Omega_{\rm b}}^{-1}$ is plotted in Fig 4, along with the best fit theoretical curve ($r_* = 0.017\,d, \nu = 2.35 \times 10^{-4}\,d^2\,\Omega_{\rm b}$). The $q = 1$ steady state profile agrees with axisymmetric theory out to a radius $r \simeq 0.24\,d$. At $r \simeq 0.26\,d$ there is a peak in the simulation curve due to a gas buildup or dense 'ridge' that runs partway around the outer edge of the disk, coinciding with Paczynski's (1977) last nonintersecting orbit. The small peak at $r \simeq 0.35\,d$ is due to the compression of the disk by the injection stream at the disk hot spot. The $r > 0.4\,d$ component of the curve is due mainly to the injection stream.

The viscosity in the simulation agrees to within 6% with Eqn 23. We found that only a small range of values for $\nu$ (i.e. $\pm 5 \times 10^{-6}\,d^2\,\Omega_{\rm b}$) gave a *reasonable* fit to the steady state disk profile. We thus have a second independent confirmation of Eqn 23 to within about 10%. Note that pressure forces were included in the simulation, which may explain some of the small discrepancy.

The value of $r_*$ giving best fit is slightly less than $r_{\rm wd}$. This is not worrying considering our open inner boundary.

The energy dissipation in the simulation is estimated using the SPH energy equation. In fact, the equation accounts for both pressure and viscous forces so in regions of expansionary flow with little shear, $\frac{{\rm d}u}{{\rm d}t}$ may be negative. Thus comparison of simulation results with Eqn 38 also yields information regarding the relative importance of pressure and viscous forces in the simulation. The radial distribution of energy dissipation in the disk at $t = 1000.00\,{\Omega_{\rm b}}^{-1}$ is shown in Fig. 5 along with Eqn 38 ($r_* = 0.017\,d$). There is very good agreement for the inner disk, revealing the predominance of viscous forces there.

## 4 RESULTS

### 4.1 Simulations

Two simulations were completed. System parameters were chosen to correspond with the eclipsing SU UMa system OY Car. Based upon the observations of Wood et al. (1989), a mass ratio $q = \frac{3}{17}$ was used. The binary orbital period $T = 0.063121$ days. A steady mass transfer rate from the secondary of $10^{-9}\,{\rm M}_\odot{\rm yr}^{-1}$ was assumed. According to the eclipse mapping observations of Rutten et al. (1992) this is a typical $\dot{M}$ for OY Car in outburst. An isothermal equation of state

$$\frac{P}{\rho} = c^2 = {\rm constant} \tag{39}$$

**Figure 6.** Particle number as a function of time for simulations 1 and 2. Using the system parameters for OY Car given in the text, an SPH particle has a mass $0.4 \times 10^{-15}$ times the binary mass. Hence this figure can be read as a disk mass versus time graph.

was used. Other calculations with different equations of state have been performed but are not detailed here. They include simulations of constant Mach number disks, and simulations of disks whose surface elements radiate as blackbodies.

With the exceptions noted below, these two simulations were completed in a very similar fashion to the nonaxisymmetric test in Section 3.4. Simulation parameters are given by Table 2. Once more th sound speed $c = 0.02\, d\, \Omega_{\rm b}$ and viscosity $\nu = 2.5 \times 10^{-4}\, d^2\, \Omega_{\rm b}$. Again the calculations were followed till $t = 1000.00\, \Omega_{\rm b}^{-1}$. Boundary conditions remain unchanged.

The two $q = \frac{3}{17}$ simulations differed only in the details of the particle injection:

**Simulation 1** Particles were added singly at $L_1$ every $\Delta t = 0.01\, {\Omega_{\rm b}}^{-1}$ with a speed $v = 0.1\, d\, \Omega_{\rm b}$ at an angle .367 rad to the binary axis in the direction of the binary rotation.

**Simulation 2** Particles were added to the calculation by placing them in a circular Keplerian orbit at the circularisation radius, $r_{\rm c}$. This is just the radius of the narrow ring about the white dwarf that forms early on in Simulation 1. For $q = \frac{3}{17}$, $r_{\rm c} = 0.1781\, d$. Thus mass and angular momentum was added to the disk at the same rate as in the first simulation.

### 4.2 Evolution

In Fig. 6, the disk mass is plotted as a function of time for the two simulations. In both cases, the mass increases to a maximum before decaying slightly to a steady value. A sticky particle simulation of Hirose and Osaki (1990) for a system with $q = 0.15$ and $\nu = 2.02 \times 10^{-4}\, d^2\, \Omega_{\rm b}$ had a similar mass-time diagram (their Fig. 5). For all three calculations the evolutionary picture is as follows. Initially the disk remains approximately axisymmetric as it increases in mass. Then the 3 : 1 resonance is encountered and the disk develops an eccentricity. In the binary frame the disk's major axis begins to rotate in a retrograde fashion with a period a few percent longer than the binary's (in the inertial frame the motion is prograde).

In these three simulations the disk became sufficiently eccentric that when the disk centre of mass passed between the two stars, the outer edge just brushed the secondary star's Roche lobe. Maximum disk mass coincided with the onset of this cyclic mass return to the secondary. The disk then finds an equilibrium between mass input from $L_1$, accretion onto the white dwarf, and mass return to the secondary. Note that in our simulations the ratio of mass accreted onto the white dwarf to mass returned to the secondary was greater than 100 : 1.

The simulation 1 disk encountered the 3 : 1 resonance at a much later time than did the simulation 2 disk. Clearly the addition of material with low specific angular momentum to the outer disk inhibited the disk's expansion to larger radii, resulting in a much larger steady state mass for simulation 1.

Figs 7 and 8 show the surface density at various epochs for simulations 1 and 2 respectively. The mass transfer stream's rôle as a 'growth inhibitor' is readily apparent. We draw particular attention to the relative states of the two disks at time $t = 50.00\, \Omega_{\rm b}^{-1}$ (top left frame). Disk 1 is confined to an almost circular region well within the tidal radius. There is a sharp drop in the density at the disk's well defined outer boundary. In contrast, disk 2 extends out to the tidal radius with a very shallow density gradient in the outer regions.

Early on in the development of the tidal instability, two armed spirals are evident in both simulations (frame 4 Fig. 7 and frame 2 Fig. 8). We identify these as the $(2\theta - 3\Omega_{\rm b} t)$ mode that is a component of the tidal instability mechanism. Later in the simulations, with several modes excited, it is harder to identify them individually.

### 4.3 Eccentricity growth rate

To determine the strength of the $l\theta - m\Omega_{\rm b} t$ mode in the disk, hereafter referred to as the $(l, m)$ mode, we follow Lubow (1991 b) and Fourier decompose the simulation particle distributions as functions of angular position and time. We choose a non-rotating coordinate system $(r, \theta)$ centred upon the white dwarf. By definition, the total strength of the $(l, m)$ mode

$$\begin{aligned} S_{l,m}(t) = (S_{\cos,\cos,l,m}(t)^2 \quad &+ \quad S_{\cos,\sin,l,m}(t)^2 \\ +S_{\sin,\sin,l,m}(t)^2 \quad &+ \quad S_{\sin,\cos,l,m}(t)^2)^{\frac{1}{2}}. \end{aligned} \qquad (40)$$

The component modes are defined as follows

$$\begin{aligned} S_{\sin,\cos,l,m}(t) \quad &= \quad \frac{2}{\pi N(1+\delta_{l,0})(1+\delta_{m,0})} \\ \int_t^{t+2\pi} \quad & \sum_{p=1}^{N} \quad \sin(l\theta_p)\cos(mt')\,{\rm d}t', \end{aligned} \qquad (41)$$

where $\theta_p$ is the angular position of particle $p$ and $N$ is the total number of particles. The integral is over a time period of $2\pi$ (one orbital period). $\delta$ is the Kronecker delta. The other component modes have similar definitions. Four disk modes were followed in the simulations:

(1,0), a measure of the disk eccentricity.

(3,3), the disk's tidal response to the $m = 3$ component of the binary potential.

**Figure 7.** Greyscale maps of the surface density in simulation 1 at times $t = 50.0, 350.0, 400.0, 450.0, 500.0, 550.0\,\Omega_{\rm b}^{-1}$. The Roche lobe and $3:1$ eccentric inner Lindblad resonance are marked in each frame. A logarithmic scale (to right) is used.

**Figure 8.** Greyscale maps of the surface density in simulation 2 at times $t = 50.0, 100.0, 150.0, 200.0, 250.0, 300.0\,\Omega_b^{-1}$. The Roche lobe and 3 : 1 eccentric inner Lindblad resonance are marked in each frame. A logarithmic scale (to right) is used.

**Figure 9.** Fourier mode analysis of disk 1 from time $t = 300.00$ to $t = 600.00\,\Omega_{\rm b}^{-1}$. This covers the interval of strongest growth in the disk eccentricity.

**Figure 10.** Fourier mode analysis of disk 2 from time $t = 0$ to $t = 200.00\,\Omega_{\rm b}^{-1}$. This covers the interval of strongest growth in the disk eccentricity

(2,3), travelling (two armed spiral) wave launched at the Lindblad resonance. This is the third mode in the TRI mechanism.

(2,2), the disk's tidal response to the $m = 2$ component of the binary potential. Lubow (1992) found that this mode cleared material out of the Lindblad resonance and interfered with the TRI mechanism.

Figs 9 and 10 plot the various mode strengths as functions of time for disks 1 and 2 respectively. In both cases the graphs cover the period from the disks' initial encounter with the resonance up till an eccentric equilibrium is reached. We draw attention to two key findings of the analysis. Firstly, in both simulations the initial growth in the eccentricity can be fit with exponential curves. Secondly the strength of the disk's $(2,2)$ tidal response diminishes over the same time interval in which the $(1,0)$ mode gains strength.

We use Eqn 3 to obtain a theoretical estimate for the eccentricity growth rate. The mean radius of the $m = 3$ eccentric inner Lindblad resonance is $r_{\rm res} = 0.455\,d$. For disk 2, the tidal instability becomes important after about $t = 50.00\,\Omega_{\rm b}^{-1}$. At this time the mean surface density at the resonance radius divided by the disk mass

$$\frac{\Sigma(r_{\rm res})}{M} = 1.15\,d^2. \tag{42}$$

The eccentricity is assumed to be independent of radius. The theoretical estimate of the eccentricity growth rate is then $\lambda_{\rm t} = 0.097$ However, given our assumptions, we feel it is probably wise to assume an uncertainty of about 10% and take

$$\lambda_{\rm t} = 0.10 \pm 0.01 \tag{43}$$

By simply plotting the (1,0) mode strength on log-linear axes and fitting a straight line to the most rapidly increasing section of the curve we obtain

$$\lambda_1 = 0.046 \pm 0.002 \tag{44}$$

for simulation 1, and

$$\lambda_2 = 0.058 \pm 0.002 \tag{45}$$

in the second case. The simulation $\lambda$ are unambiguously smaller than the theoretical value. However they are only reduced from $\lambda_t$ by a factor 2, which is surprising considering that the theory only took account of the $m = 3$ component of the binary potential and our simulations incorporate the full tidal field. In contrast, SPH simulations of Lubow (1991b) completed with the full binary potential produced only a quadratic growth in the eccentricity. We also see a significant difference in eccentricity growth rate between our two simulations. In simulation 1, the mass transfer stream is feeding mass with low specific angular momentum almost directly into the resonance region. It is to be expected therefore that this inhibits the growth of the eccentricity. It does not however disrupt the tidal instability mechanism completely.

Lubow (1992) completed a series of simulations with only the $m = 3$, and various fractions of the $m = 2$ component of the tidal potential included. He found that the tidal response generated by the $m = 2$ component was clearing material out of the $3:1$ resonance and reducing the effectiveness of the eccentric instability. In our simulations however, the $(2,2)$ mode itself is severely weakened by the disk's encounter with the $3:1$ resonance, and the tidal response's capacity to interfere with the eccentricity growth mechanism is limited.

The disks in our simulations were significantly different to those of Lubow, even before the $3:1$ resonance was encountered. We allowed the disk to build up via steady mass transfer, and by the time the disk had become resonant, the inner disk had a density versus radius profile that was close to the steady state axisymmetric solution. Lubow on the other hand initially laid down an annulus of particles on nonintersecting orbits and did not allow any form of mass addition. With viscous and pressure forces present, the annulus would spread. Lubow's disk could not approach a steady state. For the calculations he performed using only the $m = 3$ component of the tidal potential, the instability occurred immediately and grew on a time scale much shorter than the spreading time of the ring. However with the full potential included, Lubow found he had to wait much longer for the eccentric instability to be excited. To overcome that problem he gave the annulus an initial, arbitrary eccentricity.

The annuli in Lubow's calculations were hotter and less viscous than ours, but not dramatically so. He used a sound speed $c = 0.05\, d\, \Omega_{\mathrm{b}}$, two and a half times greater than ours. His viscous dissipation was characteristic of a disk with $\alpha \simeq 0.1$. Thus the kinematic viscosity $\nu$ at resonance would be about a factor three less in Lubow's simulations than in ours. However Lubow ran an $m = 3$ simulation with a kinematic viscosity increased to be somewhat closer to ours, and found no significant difference in his results.

### 4.4 Disk precession

Both disk 1 and disk 2 eventually evolved to a state of eccentric equilibrium. In this state the disks rotated in a retrograde fashion about the white dwarf primary. For simulation 1 we measured the motion of the disk centre of mass over 65 cycles to obtain a mean disk period. As a fraction of the binary period, ± 3 standard deviations

$$\frac{T_{\mathrm{d1}}}{T} = 1.08 \pm 0.02. \tag{46}$$

In simulation 2, over 118 disk cycles

$$\frac{T_{\mathrm{d2}}}{T} = 1.07 \pm 0.02. \tag{47}$$

In neither simulation was there evidence of any secular or long term trend in the disk period. The details of the disk simulations completed by various authors have already been discussed. The precession periods they obtained are listed in Table 3. It is apparent that, in addition to any dependance of $T_{\mathrm{d}}$ upon $q$, the pressure and viscosity have a significant impact upon the precession rate. This comes as a qualitative confirmation of Lubow's (1992) identification of the factors contributing to the disk's precession. In general, the superhump period is observed to decrease over the course of a superoutburst. This is perhaps indicative that the disk properties (pressure, temperature, viscosity) in the region of the resonance, change significantly as the outburst progresses.

### 4.5 Disk dissipation

Viscous dissipation releases heat in the disk. In an axisymmetric thin disk, changes in the radial disk structure occur only on the viscous diffusion time scale which is much longer than the thermal time scale. In such a disk it can be assumed that heat is radiated away at the radius at which it was generated. In our nonaxisymmetric simulations, gas in the outer regions experiences significant changes in conditions on dynamical time scales which are of similar order to thermal time scales. In nonaxisymmetric regions, we cannot assume that the disk shines brightest where the viscous dissipation is greatest. Thus in this section we do not claim to produce disk luminosity maps, we simply locate the regions of significant energy release in these precessing disks.

In Fig. 11 we plot the rate of viscous dissipation (the units are energy time$^{-1}$) for various regions of disk 2 as a function of time. The disk's total dissipation rate (uppermost curve) is dominated by a very noisy component from the inner disk. In the simulations there are relatively few particles near the inner edge where the shear forces (and dissipation) are largest. Hence the accretion of even a single particle onto the white dwarf makes a large difference to the total dissipation rate.

A second curve, showing the total dissipation rate at radii $r > 0.05\, d$ is much less noisy. Superposed on a steady signal, we see a set of evenly spaced 'humps'. The spacing of the humps matches the period of motion of the disk centre of mass. The humps are broad, each lasting approximately two and a half scaled time units, and have an irregular spiky appearance. Recall that in simulation 2 there is no mass transfer stream and hence no bright spot. In fact the stream impact with the periodically advancing and retreating disk edge (seen in simulation 1) produces another series of smaller humps that are approximately 180 degrees out of phase with those shown in Fig. 11. The first set of humps is simply caused by the periodic compression of the eccentric disk as it rotates in the white dwarf's Roche lobe. The source is an extended region in the outer disk. The best we can say is that this is not inconsistent with observational studies. For example, Naylor et al. (1987) estimated that the superhump light from OY Car came from a region that was optically thick with a temperature of approximately 8000K and an area comparable to the entire disk.

**Table 3.** Disk rotation periods obtained in eccentric disk simulations

| Author | $q$ | $\frac{T_{\mathrm{d}}}{T}$ | Comment |
|---|---|---|---|
| Whitehurst 1988a | 0.15 | 1.035 | $\nu \propto r^{\frac{1}{2}}$ |
| Lubow 1991b | 0.10 | 1.026 | $\alpha \simeq 0.1$ |
| Hirose and Osaki 1990 | 0.05 | $1.044 \pm 0.011$ | $l = 0.02$ |
| | 0.1 | $1.037 \pm 0.021$ | |
| | 0.15 | $1.060 \pm 0.011$ | |
| | 0.2 | $1.069 \pm 0.021$ | |
| | 0.25 | $1.104 \pm 0.028$ | |
| | 0.3 | $1.137 \pm 0.024$ | |
| | 0.5 | $1.226 \pm 0.021$ | |
| | 0.1 | $1.043 \pm 0.010$ | $l = 0.01$ |
| | 0.125 | $1.051 \pm 0.010$ | |
| | 0.15 | $1.060 \pm 0.011$ | |
| | 0.175 | $1.081 \pm 0.043$ | |
| | 0.2 | $1.083 \pm 0.025$ | |
| Heemskerk 1994 | 0.2 | 0.97 | |
| Murray 1995 | $\frac{3}{17}$ | $1.08 \pm 0.02$ | Simulation 1 |
| | | $1.07 \pm 0.02$ | Simulation 2 |

**Figure 11.** Rate of dissipation as a function of time for, the entire disk (upper curve), and the outer disk (lower curve) for simulation 2. Scaled units are used. The disk is in a precessing equilibrium state with an approximately constant mass (see Fig. 6).

**Figure 12.** Grey scale maps of dissipation rate in disk 2 at times t=1001.00, 1002.00, 1003.00, 1004.00, 1005.00, 1006.00 $\Omega_b^{-1}$. A logarithmic scale (shown to right) is used. The inner disk ($r < 0.05\,d$) is not shown. The Roche lobe is marked in each frame.

**Figure 13.** Dissipation rate as a function of time (in days) for simulation 2 over the period during whcih the eccentricity growth is fastest. The disk encounters the 3 : 1 resonance at approximately $t = 0.6$ days.

Fig. 12 is a sequence of six 'dissipation maps' of disk 2. The six frames are separated $1.00\,\Omega_b^{-1}$ in time and so approximately cover one full rotation of the disk about the primary. The inner portion of the disk $r < 0.05$ is not shown as the large dissipation from this region overwhelms the structure in the outer disk. A logarithmic scale is used.

Frame one shows a relatively compact disk with the dissipation dominated by the axisymmetric inner regions. Although the outer edge has a lot of structure, there is no apparent 'tail' at this stage.

In frame two ($1.00\,\Omega_b^{-1}$ later) the disk has a large tail extending out to three o'clock. This indicates that significant angular momentum and energy has been transferred to gas in the outer disk, allowing it to move to larger, more eccentric orbits. By the third frame the large tail has moved to five o'clock (remember that the motion of individual particles is anticlockwise). In frame four the tail has moved further clockwise to about seven o'clock. We note the appearance of a narrow dense, high dissipation region along the trailing edge of the tail. In the arc from about seven to nine o'clock, the outer disk experiences a negative tidal torque (see e.g. Fig. 3.6 Hirose 1991). In other words, in this region angular momentum is extracted from the outer disk and returned to the binary orbit. Thus, as the gas is braked by this tidal torque, we see a compression at the trailing edge of the tail.

By frame 5 the tail has moved closer to eight o'clock and the compression at the trailing edge has intensified. Considerable amounts of kinetic energy are being released as heat here. In between frames four and five the disk's dissipation rate has increased. Frame 6 corresponds to maximum dissipation rate. We see the tail generated over the last precession cycle in the process of being destroyed. In particular we see one of the dense knots that form in the trailing edge of the tail, falling into the axisymmetric inner disk, releasing large amounts of heat. In fact the knot shown here is responsible for the second largest dissipation spike in this particular hump.

Over the course of one rotation, the disk develops a considerable eccentric tail. This tail is then destroyed by negative tidal torques, releasing angular momentum to the binary orbit and energy as heat.

O'Donoghue (1990) attempted to use the eclipse mapping technique of Horne (1985) to obtain two dimensional maps of the superhump source. The original technique was only used to map sources that varied on time scales much longer than the eclipse time. In order to apply the technique to sources that are variable on shorter time scales, additional constraints upon the problem are needed. Hence O'Donoghue assumed that the superhump source had a fixed position in the binary frame, and that the entire superhump source varied in brightness coherently. With these assumptions, O'Donoghue considered four eclipsed superhumps in detail. In each case he found the source to be located on the edge of the disk in the approximate direction of the secondary star (i.e. towards 9 o'clock if we were looking at Fig. 12). In three eclipses the source could be resolved into three components; a compact source at the disk edge lying almost on the disk axis, and two flanking extended sources that followed the disk edge in either direction. The fourth eclipse showed a single source located near where the mass transfer stream would be expected to strike the disk edge.

There are significant differences between our two dimensional simulations, and the superhump eclipse maps of O'Donoghue. Whereas he has determined that the superhump source is located close to the binary axis, we see the dissipation occurring downstream (anticlockwise) of this location. In the simulations, the sources of excess dissipation are not stationary but move with the gas flow. Neither are they a single coherent source. It would be interesting to see how those two assumptions affected O'Donoghue's results. On the other hand, we have not treated the thermodynamics and radiation physics at all in these simulations, and we do not portray our dissipation maps (Fig. 12) as being equivalent to luminosity maps.

As a coda to this section, we show that the tidal resonance instability can generate a signal on a time scale consistent with observations. Superhumps first appear in the lightcurve of an SU UMa system within a day of it going into superoutburst. In Fig. 13 we show the dissipation rate in the outer regions ($r > 0.05$) of disk 2 as a function of time in days (recall that we chose system parameters appropriate to OY Car in superoutburst). The disk first encountered the resonance at $t \simeq 0.5$ days. The graph clearly shows a periodic signal growing to full strength within the next day or so. As we have said before, these are very simple simulations that concentrate on the dynamics of the system. The details of the disk thermodynamics and radiation have been largely ignored. That a significant periodic signal does appear on a similar time scale to superhump observations is not proof that the precessing disk model is correct. If however, the periodicity in the dissipation rate had only emerged after a much longer time, then we would suspected that the tidal instability mechanism was too weak to generate superhumps.

## 5 SUMMARY

Particle simulations of accretion disks in close binary systems have been reviewed. Particle methods are well suited to modelling highly viscous disks in gravitational potentials with little symmetry. A Smoothed Particle Hydrodynamics (SPH) scheme specifically for thin disk problems was developed. We described the use of an artificial viscosity term to provide viscous dissipation and then derived a theoretical estimate for an equivalent shear viscosity. Axisymmetric and nonaxisymmetric tests of the code confirmed the theoretical result to within 10%. This is an improvement upon earlier particle schemes for which the dissipation could only be determined empirically.

Two disk simulations for a mass ratio $q = \frac{3}{17}$, differing only in the method of mass addition to the disk, were completed. We find that the addition of material with a low specific angular momentum to the outer disk, via the mass transfer stream from $L_1$, circularises disk orbits and inhibits the disk's radial expansion.

Both our disks developed an eccentricity in a fashion consistent with the tidal resonance mechanism detailed in Lubow (1991a). The initial growth of the eccentric mode strength was exponential with a rate approximately half the predicted value. Mass transfer via the $L_1$ stream reduced the growth rate slightly. Lubow (1992) suspected that the disk's tidal response to the $m = 2$ component of the binary potential reduced the strength of the eccentric instability by removing material from the resonance region. In our simulations, the $m = 2$ tidal response weakened considerably as the disk became more eccentric.

In the binary frame, the eccentric disks rotated in a retrograde fashion with a period slightly exceeding the binary's. The rotation gave rise to a periodic excess in the rate at which energy is viscously dissipated in the disks. In the precessing disk model, this excess is ultimately the source of the superhump luminosity. We find that as the resonant disk rotates, a large eccentric tail develops. When the tail moves through a region where the tidal torques are negative it collapses in toward the centre of the disk, releasing large amounts of heat. The source of the dissipation is found to share general properties with superhump observations however the details cannot be immediately reconciled with the eclipse maps of O'Donoghue (1990). Our results are certainly inconsistent with the assumptions used to generate those maps. It is hoped that these simulations will be useful in determining new constraints for the deconvolution of superhump eclipses.

Finally we find that the periodic component of the dissipation curve developed on a time scale that was consistent with the first appearance of superhumps in a superoutburst.

*Acknowledgements* This work has been funded by an Australian Postgraduate Research Award, and by a Canadian NSERC grant. The author would like to thank Joe Monaghan for many useful discussions, and John Cannizzo, Dave Syer and Steve Lubow for their helpful remarks. Robin Humble and Christophe Pichon helped with data visulisation and MATHEMATICA respectively.